\documentclass[sigconf]{acmart}
\usepackage{amsmath,amssymb,amsfonts,bm,amsthm}
\usepackage[linesnumbered,ruled]{algorithm2e}
\usepackage{graphicx}
\usepackage[tight,normal]{subfigure}
\usepackage{multirow,makecell,url,footmisc}
\usepackage{enumerate}
\usepackage{lineno}
\usepackage{textcomp}
\usepackage{booktabs} 

\DeclareMathOperator*{\argmax}{argmax}

\copyrightyear{2017} 
\acmYear{2017} 
\setcopyright{acmcopyright}
\acmConference{KDD'17}{}{August 13--17, 2017, Halifax, NS, Canada.}
\acmPrice{15.00}
\acmDOI{http://dx.doi.org/10.1145/3097983.3098134}
\acmISBN{978-1-4503-4887-4/17/08}

\begin{document}

\title{Optimized Cost per Click in Taobao Display Advertising}

\author{Han Zhu, Junqi Jin, Chang Tan, Fei Pan, Yifan Zeng, Han Li, Kun Gai}
\affiliation{%
  \institution{Alibaba Group}
}
\email{{zhuhan.zh, junqi.jjq, tanchang.tc, pf88537, yifan.zyf, lihan.lh, jingshi.gk}@alibaba-inc.com}

\begin{abstract}

Taobao, as the largest online retail platform in the world, provides billions of online display advertising impressions for millions of advertisers every day. For commercial purposes, the advertisers bid for specific spots and target crowds to compete for business traffic. The platform chooses the most suitable ads to display in tens of milliseconds. Common pricing methods include cost per mille (CPM) and cost per click (CPC). Traditional advertising systems target certain traits of users and ad placements with fixed bids, essentially regarded as coarse-grained matching of bid and traffic quality. However, the fixed bids set by the advertisers competing for different quality requests cannot fully optimize the advertisers' key requirements. Moreover, the platform has to be responsible for the business revenue and user experience. Thus, we proposed a bid optimizing strategy called optimized cost per click (OCPC) which automatically adjusts the bid to achieve finer matching of bid and traffic quality of page view (PV) request granularity. Our approach optimizes advertisers' demands, platform business revenue and user experience and as a whole improves traffic allocation efficiency. We have validated our approach in Taobao display advertising system in production. The online A/B test shows our algorithm yields substantially better results than previous fixed bid manner.

\end{abstract}

%
%



\maketitle

\section{Introduction}\label{section:Introduction}

Advertising fosters the rise of new brands and keeps existing quality brands youth forever. Online advertising \cite{goldfarb2011online,lahaie2007sponsored,evans2009online,karande2013optimizing}, a marketing strategy involving the use of the internet as a medium to obtain website traffic and target, and deliver marketing messages to the right customers, has experienced an exponential increase in the growth since the early 1990s. Real-time bidding (RTB) \cite{perlich2012bid,yuan2014empirical,muthukrishnan2010data} technology in online advertising allows advertisers to bid for every individual impression. And lots of research \cite{zhang2014optimal,yuan2014survey,yuan2013real,zhang2016bid} has found effective and efficient bidding strategies to maximize unilateral economic surplus of a party, such as advertisers, consumers and intermediary platforms.

More than RTB systems, Taobao, called ``the country's biggest online marketplace'' by the Economist \cite{econ2015omist}, established one of the most advanced online advertising system in the world. In both mobile app and PC website of Taobao, selected ads are presented to users in specific spots. In this paper, we focus on the bid optimization problem of the indispensable CPC display advertising in Taobao mobile app. Two ad formats involved are as follows:


\begin{figure}[tb]
  \centering
  \includegraphics[width=0.75\columnwidth]{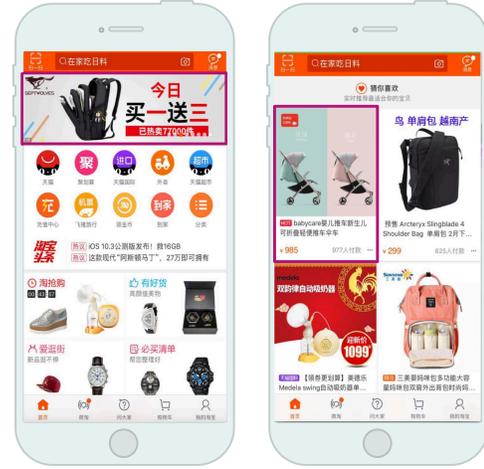}
  \caption{Banner and item CPC ads displayed on Taobao mobile app home.}
  \label{fig:banner_item}
\end{figure}

\begin{itemize}
\item \emph{Banner CPC Ads:} The ads appear in the top banner of Taobao home page as Figure \ref{fig:banner_item}. Advertisers set up campaigns for a single item, a store or a brand.

\item \emph{Item CPC Ads:} Single items are displayed to users in the \emph{Guess What You Like} column including about two hundred spots, three of which are for advertising and the others are for recommendation as Figure \ref{fig:banner_item}.
\end{itemize}

Connecting users and advertisers, Taobao advertising platform forms its own unique ecosystem characterized by:
\begin{itemize}
\item First, unlike most RTB systems, for which it's difficult to obtain complete user data, Taobao itself acts as demand side and supply side at the same time. This ecologically closed-loop system enables Taobao to collect integrated user data and ad campaigns' information.

\item Second, most advertisers in the system are small and medium-sized ones who are more concerned about the increase in revenue than promoting their brands. Therefore the increase in gross merchandise volume (GMV) can better benefit these advertisers.

\item Third, while different advertisers could pursue different key performance indicators (KPI, e.g., impressions, clicks, conversions or return of investment (ROI)), they bid for clicks on Taobao platform, i.e., CPC is adopted. We will discuss other methods such as cost per mille (CPM) and cost per sale (CPS) later.

\item The last but the most important is that advertising spots must meet the media requirements, which is measured by indices such as click-through rate (CTR), conversion rate (CVR), GMV, etc. Here is an example of GMV analysis. First, we hope that the introduction of business traffic does not unduly affect user experience. Thus, setting GMV requirements achieves a win-win situation of business revenue and user experience. Second, as Taobao's advertisers are precisely Taobao's sellers, with sellers using an approximately fixed percentage (taking rate) of revenue for marketing purposes, raising GMV will result in advertisers increasing their advertising budgets, which brings long-term benefits to the platform.
\end{itemize}

Weighing the pros and cons, we adopt CPC in the two ad formats. Although advertisers assume less risk with CPS \cite{edelman2007internet,aggarwal2006truthful,varian2007position}, compared to CPC, CPS ignores the value of clicks, providing worse traffic liquidating efficiency. Since the involved ad formats are mainly for small and medium-sized advertisers, CPM poses higher risk while CPC allows advertisers to control the cost of clicks and the platform takes the risk of turning page views to clicks. With Taobao's complete data ecology, standardized e-commerce advertising and interactive process, CPC is sufficiently effective.

Many state-of-the-art systems such as Facebook's \cite{ocpm2012} use different designs from Taobao. To some large social networking services (SNSs), for example, through optimized cost per mille (oCPM), advertisers can bid for click and actually pay per impression \cite{ocpm2012}. SNS advertising interactions are usually divergent, such as like, click, share, etc., while Taobao's transactions are often accomplished by simple serial clicks. From the data ecology point of view, after ad clicks, Taobao users' all behaviors are still on Taobao platform, which provides conditions for follow-up interaction-based deductions. However, the SNS usually lets advertisers bid for clicks or other actions and then converts to equivalent CPM manner, which in mechanism encourages advertiser to upload real follow-up interaction data and further optimizes the bid.

In previously mentioned two ad formats, taking into account the ecology, efficiency, etc., we choose CPC method which is the focus of this paper.

Taobao's advertising system includes filtering millions of ads and ranking of these candidate ads. First, mining user preference inferred from its behavior data and the ad item's details, Taobao targeting system \cite{provost2009audience,raeder2012design} trains models to filter mass amount of ads for each page view (PV) request, which is called \emph{matching} stage. Different from the recommendation \cite{schafer2007collaborative} not involving advertisers, the matching service recalling related users has to reflect the advertisers' bidding will and ensuring market depth. Secondly, real-time prediction (RTP) engine predicts click-through rate (pCTR) for each eligible ad. Thirdly, traditionally, these candidate ads are ranked by $bid * pctr$ and displayed based on the order to maximize effective cost per mille (eCPM sorting mechanism).

Advertisers always expect the bid to match the traffic quality. Due to technical limitations, traditional methods can only set fixed bids for specific user groups and ad slots for coarse-grained traffic differentiation, however, advertisers are looking for further fine-grained matching of bids and traffic quality. Ranking process based on the fixed bids has two defects. On the one hand, it is inefficient that a fixed bid set by an advertiser deals with continuous internet traffic of different commercial qualities; on the other hand, traditional methods maximize eCPM to pursue short-term commercial revenue, however, can not optimize and control media requirements such as GMV, detrimental to Taobao's long-term interests.

For these two issues, from the perspective of advertisers, oCPM in some SNSs \cite{ocpm2012} converted equivalently from other bidding objectives, is able to maximize advertisers' interests, however, may not guarantee the platform ecological health such as GMV; from another aspect, excessive pursuit of media requirements like GMV by modifying the ranking formula $bid * pctr$ can not bring effective commercial benefits to advertisers and the platform.

In order to solve above problems, we propose optimized cost per click (OCPC) with following characteristics: for each PV request, on the premise of optimizing the advertiser's demands, OCPC adjusts the bid toward the true value of the traffic quality, and meanwhile maximizes a composite score reflecting overall ecology of user experience, advertisers' interests, and platform's revenue, by keeping eCPM sorting mechanism unchanged; our design allows us to adapt the OCPC system flexibly with lower costs based on the changing needs of the business. We expect through optimizing the traffic matching efficiency, our OCPC achieves a comprehensive upgrade of all the user, advertiser and platform indices. It's worth mentioning that enhanced cost per click \cite{ecpc} (ECPC) in Google AdWords also attempts to adjust the bid according to the potential conversion rate. However, besides conversion rate, platform indices like GMV, which are crucial elements for Taobao platform, cannot be optimized directly in ECPC manner.

Our major contributions are summarized below. (i) We illuminate some characteristics of Taobao display advertising system and its subsystems. (ii) We propose a novel bid optimization approach which achieves the overall optimization of advertisers' interests, user experience and platform revenue of Taobao ecology. (iii) Comprehensive offline and online experiments are conducted to verify the effectiveness of the proposed OCPC mechanism.

The rest of this paper is organized as follows: Section \ref{section:Flow} gives a brief introduction to Taobao advertising system. Section \ref{section:Mechanism} presents the OCPC details. Section \ref{section:Model} introduces the prediction process. At last, Section \ref{section:Experiment} focuses on the experimental results about the proposed approach, including model effectiveness estimation, offline experimental mechanism, and online A/B test performance.

\section{System Architecture} \label{section:Flow}
\begin{figure}[tb]
  \centering
  \includegraphics[width=0.8\columnwidth]{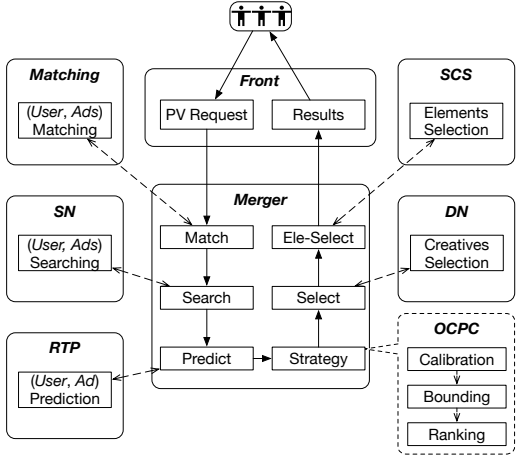}
  \caption{The star schema of Taobao display advertising system and the proposed bid optimization strategy used in it.}
  \label{fig:architechture}
\end{figure}

This section describes how data and information flow in Taobao's display ads system as Figure \ref{fig:architechture}, which is essential to help understand why and how bid optimization works. Each system component and the sequence of events handled in them from the foremost page view request to the ultimate impression are highlighted as follow:

\textbf{Front Server} receives a page view request from a user and hand out to \textbf{Merger Server} which acting as a central coordinator communicates with other components during the whole process. Merger Server requests \textbf{Matching Server} to analyze the user and get a list of feature tags according to the advertisers' user targeting requirements. Through Merger Server, these tags are delivered to \textbf{Search Node (SN) Server} for searching particular candidate ads along with the bids. In aforementioned \emph{Guess What You Like}, the number of candidates is reduced from thousands to about four hundreds. Then, \textbf{Real-time Prediction (RTP) Server} predicts the click-through rate (pCTR) and conversion rate (pCVR) for the candidates from SN. In terms of CTR prediction \cite{chen2016deep,graepel2010web,he2014practical}, we use mixture of logistic regression (MLR, which is also called as LS-PLM \cite{gai2017learning}) model to deal with particular high dimensional, i.e., usually hundreds of millions of dimensions, sparse and binarized features. As a part of merger, \textbf{Strategy Layer} contains the main logic of OCPC which optimizes traffic allocation by ranking stage based on pCTR, pCVR and bid. The strategy layer is also responsible for the follow-up ads duplication removal, and final impression price calculation under generalized second-price auction (GSP). According to the rank of ads, titles and image addresses are extracted by \textbf{Data Node (DN) Server}, which are further optimized by \textbf{Smart Creative Service (SCS)}. At last, the front server returns ad results to the mobile app or PC website. And the subsequent click or conversion will be recorded in the log system. All subsystems together constitute a complete data ecology based on which we introduce OCPC strategy in the next section.

\section{Optimized Cost per Click} \label{section:Mechanism}

In this part, we first mathematically describe the demands of advertisers and conditions for optimization. Secondly, we propose an algorithm to optimize the platform ecology index and platform revenue. At last, relevant details are introduced. Practically, our algorithm framework applies to a wide range of advertisers' demands and platform ecology indices, such as number of page views, clicks, conversions, etc. As a typical case, this paper sets \textbf{ROI and gaining quality traffic} as the advertisers' demand, and \textbf{GMV} as the platform ecology index, which along with \textbf{platform revenue} are optimized by adjusting the advertisers' bids. Suppose $A$ is the set of ad campaigns that are eligible for a PV request. With this specific PV request, for each campaign $a \in A$, there exists a preseted corresponding bid $b_a$ by the advertiser. For each $b_a$, the role of OCPC algorithm is to adjust it and find an optimized $b_a^*$ to achieve the pre-designated various optimization requirements.

\subsection{Optimization Scope} \label{section:optScope}

\paragraph{\textbf{ROI Constraint}}
Taking into account the small and medium-sized advertisers being more concerned about the marketing effect, we choose to optimize their revenue (GMV) while keeping or improving ROI as a primary application of our algorithm. Here we introduce relevant notations and finally derive the mathematical representation of ROI.

First, we define the probability of transaction conversion $c$ conditioned on a user $u$ and a clicked ad $a$ as $p(c|u, a)$. For a specific item, note that in $p(c|u, a)$ the ad spot is not considered as a condition for different spots eventually leading to the same item page. For a particular ad campaign $a$, define $v_a$ as the predicted pay-per-buy (PPB) by consumers, i.e., the seller's revenue. Thus, the expected GMV for a single click is $p(c|u, a) * v_a$.

Although the actual cost is calculated according to GSP mechanism, here we suppose the cost of a click paid by the advertiser is $b_a$. So the expected ROI for a single click is derived as Eq.\ref{eq:single_roi}.
\begin{equation}
roi_{(u, a)} = \frac{p(c|u, a) * v_a}{b_a} \label{eq:single_roi}
\end{equation}

Further, the overall ROI of ad $a$ across different users and clicks is derived as Eq.\ref{Formula:roi}, where $n_u$ is the total number of clicks for a user over a period of time. (We suppose the ROI is for a particular crowd and a spot, thus $b_a$ is consistent.)
\begin{equation}
roi_{a} = \frac{v_a\cdot\sum_u n_u\cdot p(c|u, a)}{b_a \cdot\sum_u n_u} = \frac{E_u[p(c|u, a)] \cdot v_a}{b_a} \label{Formula:roi}
\end{equation}

Equation \ref{Formula:roi} indicates that the advertiser's overall ROI is determined by three factors: the expectation of conversion rate $E_u[p(c|u, a)]$, the predicted $v_a$ and bid $b_a$, among which $v_a$ is inherent for each ad, and $E_u[p(c|u, a)]$ is regarded as stationary in each particular auction.

In practice, the current prediction model is used to predict pCVRs of competing ads from past few days and the largest, smallest $10\%$ of these CVRs are eliminated, with average of the remaining composing current $E_u[p(c|u, a)]$. \textbf{The goal of bid optimization requires that $roi_a$ should keep unchanged or be improved (so called ROI constraint), and advertisers can gain more high quality traffic.}

\begin{figure}[tb]
  \centering
  \includegraphics[width=0.55\columnwidth]{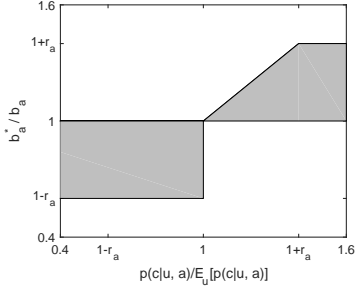}
  \caption{The bid optimization scope (the gray area) under ROI constraint.}
  \label{fig:roi}
\end{figure}

\paragraph{\textbf{Bid Optimization Boundary}}
Equation \ref{Formula:roi} proves the linear relationship between $roi_a$ and $E_u[p(c|u, a)]$, i.e., bid optimization that satisfying $\frac{b_a^*}{b_a} \leq \frac{p(c|u, a)}{E_u[p(c|u, a)]}$ will prevent ROI from falling. Along with considering advertisers' demands of gain quality traffics, we conduct the following bid optimization principles: raise the bid under ROI constraint to help advertisers compete for quality traffics ($\frac{p(c|u, a)}{E_u[p(c|u, a)]} \geq 1$), and depress the bid to save cost for those low quality traffics ($\frac{p(c|u, a)}{E_u[p(c|u, a)]} < 1$). The bid optimization range that compromised quality and quantity is illustrated in the gray area of Figure \ref{fig:roi}, based on the ratio of $p(c|u, a)$ and $E_u[p(c|u, a)]$. Note that there exists a fixed threshold $r_a$ (e.g., $40\%$), for the sake of safety and business settings. The lower bound is essential to avoid the situation that some advertisers may get little traffic when optimizing their ROI.

With the area depicted in Figure \ref{fig:roi}, the lower and upper bounds, denoted as $l(b_a^*)$ and $u(b_a^*)$, of bid optimization for an ad campaign $a$ are as Eq. \ref{Formula:lower_bound} and \ref{Formula:upper_bound}. It's worth to emphasize that the bid optimization boundaries can be generalized to refer to other pursuits of advertisers, not limited to ROI. If bid optimization is not authorized by some advertisers, the corresponding lower and upper bounds both equal to $b_a$.
\begin{equation}
\label{Formula:lower_bound}
\scriptsize
l(b_a^*) = \left\{
   \begin{aligned}
   &b_a \cdot (1 - r_a), & \frac{p(c|u, a)}{E_u[p(c|u, a)]} < 1\\
   &b_a, & \frac{p(c|u, a)}{E_u[p(c|u, a)]} \geq 1 \\
   \end{aligned}
   \right.
\end{equation}
\begin{equation}
\label{Formula:upper_bound}
\scriptsize
u(b_a^*) = \left\{
   \begin{aligned}
   &b_a, & \frac{p(c|u, a)}{E_u[p(c|u, a)]} < 1\\
   &b_a \cdot \min (1 + r_a, \frac{p(c|u, a)}{E_u[p(c|u, a)]}), & \frac{p(c|u, a)}{E_u[p(c|u, a)]} \geq 1 \\
   \end{aligned}
   \right.
\end{equation}

\subsection{Ranking}

Optimizing bid price in the given boundary can help advertisers gain better quality traffics and higher ROI. However, different bid price $b_a^*$ chosen from the feasible region might result in different ad ranks under eCPM sorting mechanism (i.e., ads are still ranked by $pctr * bid$ after bid optimization), and consequently bring different revenues or other indicators. In the following content of this section, we'll introduce our novel way to choose $b_a^*$ from the feasible region, which can attain best composite index that has taken pursuits from all sides into account, on the premise of holding eCPM sorting mechanism.

Assuming that we're going to display one ad under eCPM sorting mechanism, we expect the ad to maximize the following objective
\begin{align}
\max_{b_1^*,\cdots,b_n^*} & \hspace{8pt} f(b_k^*) \label{Formula:objective} \\
s.t. & \hspace{8pt} k = \mathop{\argmax}_{i} \hspace{5pt} pctr_i * b_i^* \label{Formula:obj1} \\
      & \hspace{8pt} l(b_i^*) \leq b_i^* \leq u(b_i^*), i = 1, \cdots, n \label{Formula:obj2}
\end{align}
where $n$ is the number of eligible ads in a PV, i.e., $\|A\|$, and $f(\cdot)$ is the function that can give a composite index which has included pursuits from all sides. Without loss of generality, we assume that $f(b_i^*)$ is monotone increasing w.r.t. $b_i^*$. Condition in Eq. \ref{Formula:obj1} means that the auction-winning ad is the top ranked $k$th ad under eCPM sorting, and the above optimization problem is to maximize $f(b_k^*)$ of the auction-winning ad. Condition in Eq. \ref{Formula:obj2} ensures the optimized bid price in the determined scope. There are two meanings for the optimization problem presented in Eq. \ref{Formula:objective}. On the one hand,  we attempt to select the $k$th ad that could have the largest $f(b_k^*)$ value; on the other hand, the bid price of each ad $i \in A$ should be adjusted to make sure that the selected $k$th ad can have the largest eCPM. For $f(\cdot)$, we give the following two examples
\begin{align}
\label{Formula:sortExample}
   & f_1(b_k^*) = pctr_k * pcvr_k * v_k,\notag \\
   & f_2(b_k^*) = pctr_k * pcvr_k * v_k + \alpha * pctr_k * b_k^* \notag
\end{align}
where $f_1(\cdot)$ tends to prompt  Taobao platform's overall GMV, which is the revenue of all advertisers. And $f_2(\cdot)$ is a compromise of Taobao's GMV and advertising revenue. Note that $\alpha$ is the trade off coefficient between GMV and advertising revenue, and different $\alpha$ value could result in different goal of bid optimization, just in the way presented in Eq. \ref{Formula:objective}.

The remaining work of ranking is to find $b_a^*$ for each $a$ that can maximize the objective in Eq. \ref{Formula:objective}. Analogy to the boundary of optimized bid price, we derive the boundary of $pctr_a * b_a^*$ as $l(s_a^*)$ and $u(s_a^*)$, called the lower and upper bound of optimized rank score $s_a^*$ ($l(s_a^*) = pctr_a * l(b_a^*)$, $u(s_a^*) = pctr_a * u(b_a^*)$, according to Eq. \ref{Formula:lower_bound} and \ref{Formula:upper_bound}). To optimize Objective \ref{Formula:objective}, we just need to sort ads in descending order of $f(u(b_i^*))$ (note that we use bid's upper bound $u(b_i^*)$ here because we assume $f(b_i^*)$ is monotone increasing w.r.t. $b_i^*$) for each $i\in A$, then choose the first ad $k$ whose $u(s_k^*)$ is no less than all other ads' $l(s_i^*)$ (to make sure that Constraints \ref{Formula:obj1} and \ref{Formula:obj2} can be satisfied) as the result to display and set $b_k^* = u(b_k^*)$. Last, update bid prices for other candidates in their feasible region, which ensure ad $k$ has the largest eCPM. 


Returning to the real scenario that there might be more than one (e.g., $N$) ads displayed in each PV, we propose a greedy algorithm in Algorithm \ref{algorithm:reranking} and give a brief explanation as follow.

First, we sort ads according to $f(\cdot)$ (line 3) and pick an ad out (derive the ad as $k$, lines 4-5) by optimizing the objective function in Eq. \ref{Formula:objective}. Then, we update remaining ads' $u(s_i^*)$ by limiting them to no more than $u(s_k^*)$ (correspondingly update $u(b_i^*)$ to ensure that ad $k$ could have the largest eCPM after bid optimization, as Constraint \ref{Formula:obj1}, lines 8-11). Afterwards, we repeat the above two steps until all $N$ ads are picked out (lines 2-12). Last, set $b_i^*$ of all ads to their bid price upper bound $u(b_i^*)$ (lines 13-15). The time complexity of the proposed ranking algorithm is $O(N * \|A\|* \log\|A\|)$. Typically, $N$ is a small number (e.g., $N = 3$ in \emph{Item CPC Ads}) that real-time response won't be an issue.

\begin{example}
Here we give an example to help understand the ranking algorithm. Suppose that there are $4$ eligible ads in $A$ given in Table \ref{table:example}, and the number of ads to display is $N = 2$. Now, we are going to select $2$ ads from \textit{Ad} 1-4 in Table \ref{table:example}. According to the proposed ranking algorithm, these 4 ads are sorted in descending order of $f_2(u(b^*))$. The largest rank score lower bound is $l(s_3^*) = 0.09$ (marked in blue in Table \ref{table:example}). And the top ranked ad's $u(s_1^*) = 0.112$ (also marked in blue),  which is larger than $0.09$. Thus, \textit{Ad} 1 is picked out from $A$ and inserted into the winning set $\mathcal{A}$, and the candidate set $A$ is updated to Table \ref{table:example1}, according to lines 6-11 in Algorithm \ref{algorithm:reranking} (updated cells are in red). Afterwards, \textit{Ad} 3 rather than \textit{Ad} 2 is selected as another winning ad in the second cycle, because \textit{Ad} 2's rank score upper bound $u(s_2^*) = 0.075$, which is smaller than $0.09$. Then, the loop ends because $\|\mathcal{A}\| == N$. Finally, the bid optimization result of each ad is given in Table \ref{table:example2}. 
\end{example}

\begin{table}[!hbp]
\small
\begin{tabular}{|c|c|c|c|c|c|c|}
\hline
Ad \# & $pctr$ & $bid$ & $u(b^*)$ & $u(s^*)$ $10^{-2}$ & $l(s^*)$ $10^{-2}$  & $f_2(u(b^*))$ \\
\hline
1 & 0.04 & 2 & 2.8 & {\color{blue}11.2} & 8  & 0.312 \\
\hline
2 & 0.05 & 1.5 & 1.5 & 7.5 & 4.5 & 0.255 \\
\hline
3 & 0.06 & 1.5 & 1.95& 11.7 & {\color{blue}9}  & 0.237 \\
\hline
4 & 0.04 & 1 & 1 & 4 & 3.6 & 0.14 \\
\hline 
\end{tabular}
\caption{$4$ eligible ads in $A$ and their pCTRs, bids, etc., and $\alpha$ in $f_2(\cdot)$ is set to $1$. The upper bound of each ad's rank score $u(s^*) = pctr * u(b^*)$, and the lower bound of each ad's rank score $l(s^*) = pctr * l(b^*)$.}
\label{table:example}
\end{table}

\begin{table}[!hbp]
\small
\begin{tabular}{|c|c|c|c|c|c|c|}
\hline
Ad \# & $pctr$ & $bid$ & $u(b^*)$ & $u(s^*)$ $10^{-2}$ & $l(s^*)$ $10^{-2}$  & $f_2(u(b^*))$ \\
\hline
2 & 0.05 & 1.5 & 1.5 & 7.5 & 4.5 & 0.255 \\
\hline
3 & 0.06 & 1.5 & {\color{red}1.86} & {\color{red}11.2} & {\color{blue}9}  & {\color{red}0.232} \\
\hline
4 & 0.04 & 1 & 1 & 4 & 3.6 & 0.14 \\
\hline 
\end{tabular}
\caption{Remained $3$ ads and their updates in $A$ after \textit{Ad} 1 is picked out. Updated cells are in red.}
\label{table:example1}
\end{table}

\begin{table}[!hbp]
\small
\begin{tabular}{|c|c|c|c|c|c|c|}
\hline
Ad \# & $pctr$ & $bid$ & $b^*$ & $f_2(u(b^*))$ & eCPM  \\
\hline
1 & 0.04 & 2 & 2.8 & 0.312 & 0.112 \\
\hline
3 & 0.06 & 1.5 & 1.86 & 0.232 & 0.112 \\
\hline
2 & 0.05 & 1.5 & 1.5 & 0.255 & 0.075 \\
\hline
4 & 0.04 & 1 & 1 & 0.14 & 0.04 \\
\hline 
\end{tabular}
\caption{Bid optimizatoin result of each eligible ad.}
\label{table:example2}
\end{table}

By such ranking strategy, we decouple the final sorting index and the goal of advertising traffic. On the one hand, ads can still be sorted by $pctr * bid$, which is the way to maximize eCPM; on the other hand, the advertising platform can choose ads according to other pursuits by different $f(\cdot)$. Another concerned problem is about budget constraint of advertisers. Once an ad campaign spends out its budget, it will be excluded from the following auctions, which would not affect the bid optimization process.

\begin{algorithm} 
\caption{Ranking Algorithm}
\label{algorithm:reranking}
\KwIn{Ad list $A$, corresponding boundaries of bid price} 
\KwOut{Optimized bid prices $b_a^*$ for $\forall a \in A$} 
Winning set $\mathcal{A} = \emptyset$\;
\Repeat{$\|\mathcal{A}\| == N$ or $A == \emptyset$}
{
	Sort ads in $A$ in descending order of $f(u(b_i^*))$\;
	$t \gets $ the largest $l(s_a^*)$ for $\forall a \in A$\;
	Find the first ad $k$ from $A$ that $u(s_k^*) \geq t$\;
	$\mathcal{A} = \mathcal{A} \cup \{k\}$\;
	$A = A \backslash \{k\}$\;
	\For{$i \in A$}{
		$u(s_i^*) = \min (u(s_i^*), u(s_k^*))$\;
		$u(b_i^*) = \min (u(b_i^*), \frac{u(s_i^*)}{pctr_i})$\;
	}
}

\For{$i \in \mathcal{A} \cup A$}{
	$b_i^* = \frac{u(s_i^*)}{pctr_i}$\;
}

Return $b_a^*$ for each ad in $\mathcal{A} \cup A$\; 
\end{algorithm}

\subsection{Algorithm Details}

After introducing the core ideas in our OCPC mechanism, we'll going to detail the whole strategy layer.

\paragraph{\textbf{Calibration}}
From the historical experience of maintaining the advertising system, we find that inherent bias exists on the predicted values used in OCPC layer, which could affect the algorithm effectiveness. Since it's difficult to do adjustment in model training, we do calibration after prediction at the beginning of OCPC layer. 

\begin{figure}[h]
    \centering
    \subfigure[Before calibration]{
        \includegraphics[width=0.22\textwidth]{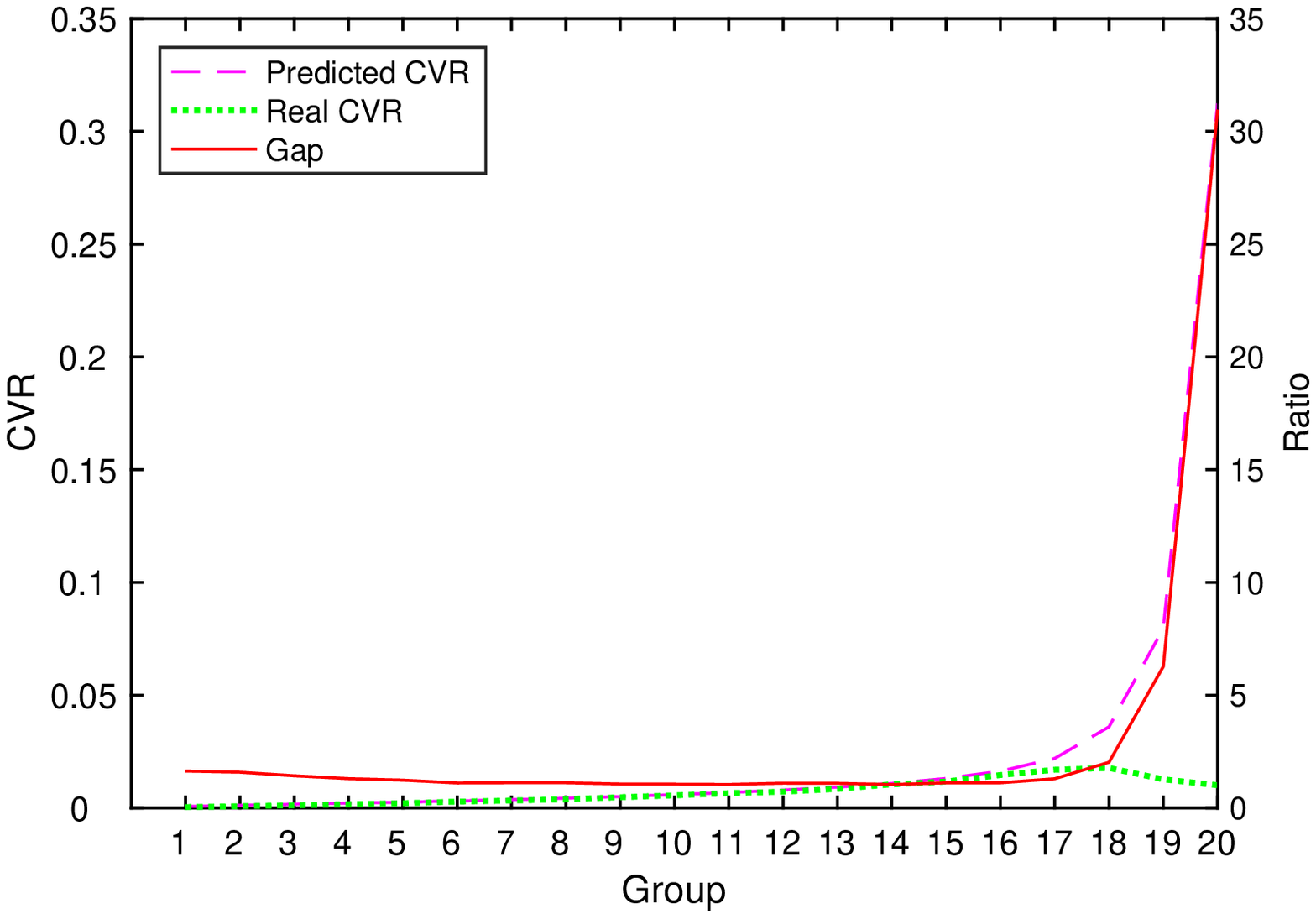}
        \label{fig:cvr_qcurve_before}
    }
    \subfigure[After calibration]{
        \includegraphics[width=0.22\textwidth]{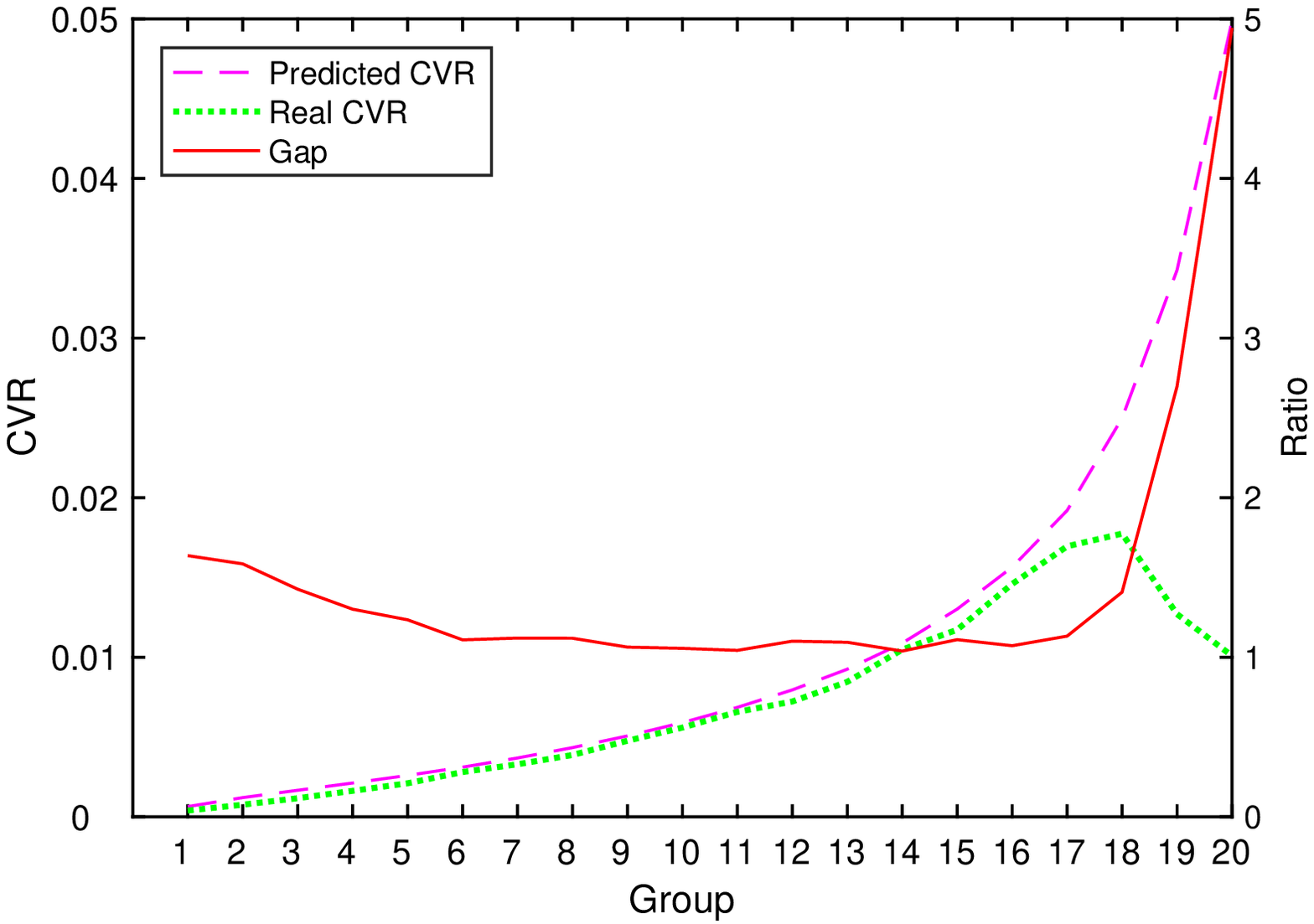}
        \label{fig:cvr_qcurve_after}
    }
    \caption{The gap between predicted and real CVR w.r.t. different pCVR level before and after calibration ($tc = 0.012$, from Jan 10, 2017 to Jan 16, 2017).}
    \label{fig:cvr_qcurve}
\end{figure}

We'll take pCVR calibration as the example. The RTP module usually gives a larger estimated CVR value when the actual CVR is in a high level. This phenomenon is illustrated in Figure \ref{fig:cvr_qcurve}. We divide all ads to 20 groups according to their pCVR. The corresponding real CVR and the ratio of predicted and real CVR are draw in the figure. We can see that the ratio becomes larger in groups with large pCVR. Thus, We calibrate the predicted CVR as
\begin{equation}
\label{equ:calibration}
\scriptsize
p(c|u, a) = 
\left\{
   \begin{aligned}
   &p(c|u, a),& p(c|u, a) < tc \\
   &tc * (1 + \log (\frac{p(c|u, a)}{tc})), & p(c|u, a) \geq tc \\
   \end{aligned}
   \right.
\end{equation}
where $tc$ is the calibration threshold, typically $0.012$ in practice. Those pCVR values that are larger than $tc$ will be calibrated with Eq. \ref{equ:calibration}, which is an intuitive way that aims to reduce the gap between predicted and real CVR for ads with large pCVR values. After calibration, we can see from Figure \ref{fig:cvr_qcurve} that the gap drops significantly in high pCVR region.

\paragraph{\textbf{Overall OCPC Strategy}}
In Algorithm \ref{algorithm:overall}, we give the overview of the over OCPC strategy, from calibration to ranking. Lines 1-4 with functions $calibrate$ and $calculateBoundary$ have linear time complexity $O(\|A\|)$. The time complexity of $rank$ function is $O(N * \|A\|* \log\|A\|)$. Therefore, the run time performance bottleneck of OCPC strategy is the ranking stage. Considering the typical value of $\|A\|$ (about hundreds) and $N$, the real-time performance is not an issue for the proposed approach.
\begin{algorithm} 
\caption{OCPC Algorithm}
\label{algorithm:overall}
\KwIn{Eligible ads $A$, and corresponding predictions}
\KwOut{Optimized bid price $b_a^*$ for $a \in A$} 
\For{$i \in A$}
{
	calibrate()\;
	calculateBoundary()\; 
}
rank()\;
return each $b_a^*$\; 
\end{algorithm}

\section{Model Estimation} \label{section:Model}
The stated bid optimization boundary of OCPC is extremely dependent on CVR prediction. Meanwhile, other predicted values like pCTR will also affect the performance of proposed strategy for the most part. In this section, we are going to focus on the prediction models, along with the accuracy and stability of predicted values.

\subsection{Model and Features} \label{section:Prediction}
In Taobao estimation, we have features of user and campaign which are sparse and have tens of millions dimensions. Logistic Regression is a widely used algorithm in tasks like CTR prediction \cite{richardson2007predicting}. However, the problem to solve may be non-linear. Therefore, we use mixture of logistic regression (MLR, which is also called as LS-PLM \cite{gai2017learning}) algorithm in RTP server. We do not expand more about MLR here, instead, we are going to introduce the composition of feature to help understand how the learning model work.
\begin{figure}[tb]
  \centering
  \includegraphics[width=0.95\columnwidth]{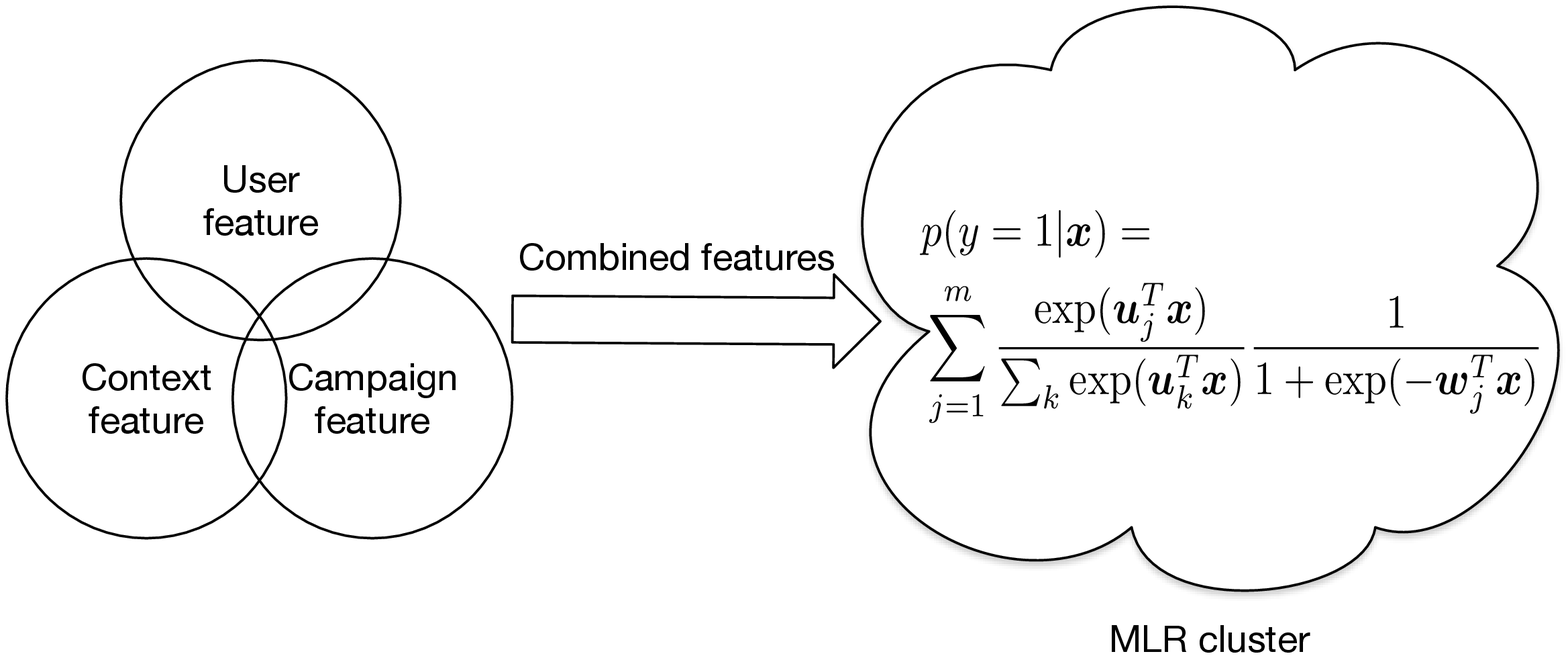}
  \caption{Feature composition of CTR and CVR prediction model.}
  \label{fig:feature}
\end{figure}

In Figure \ref{fig:feature}, we illustrate the feature composition in CTR and CVR prediction. We'll give a brief introduction to these three kinds of features along with their combinations. Context feature is features related to the context. For example, spot position feature (we call it PID feature) is used to distinguish different spots (e.g., spots in Android or IOS). User feature mainly contains user profile features (like gender, age, etc.) and user behavior features (e.g., click times of different categories in a period of time). Campaign feature is consists of features like ad ID. Beside separate features in those three kinds, their combinations (e.g., the Cartesian Product of nick name and ad ID) are also used. Furthermore, In CVR prediction, the results of click quality model (used to qualify a click behavior) are used as input, which has shown significant improvement in practice.

In CTR model, positive samples are collected from those clicked impressions. And the negative samples are those not clicked impressions. In CVR model, positive samples are those clicked and converted impressions, and the negative samples are those clicked but not converted impressions. New models are trained every day to eliminate the variance between different days.

\subsection{Model Performance} \label{section:Performance}
\begin{figure}[h]
    \centering
    \subfigure[CTR model]{
        \includegraphics[width=0.22\textwidth]{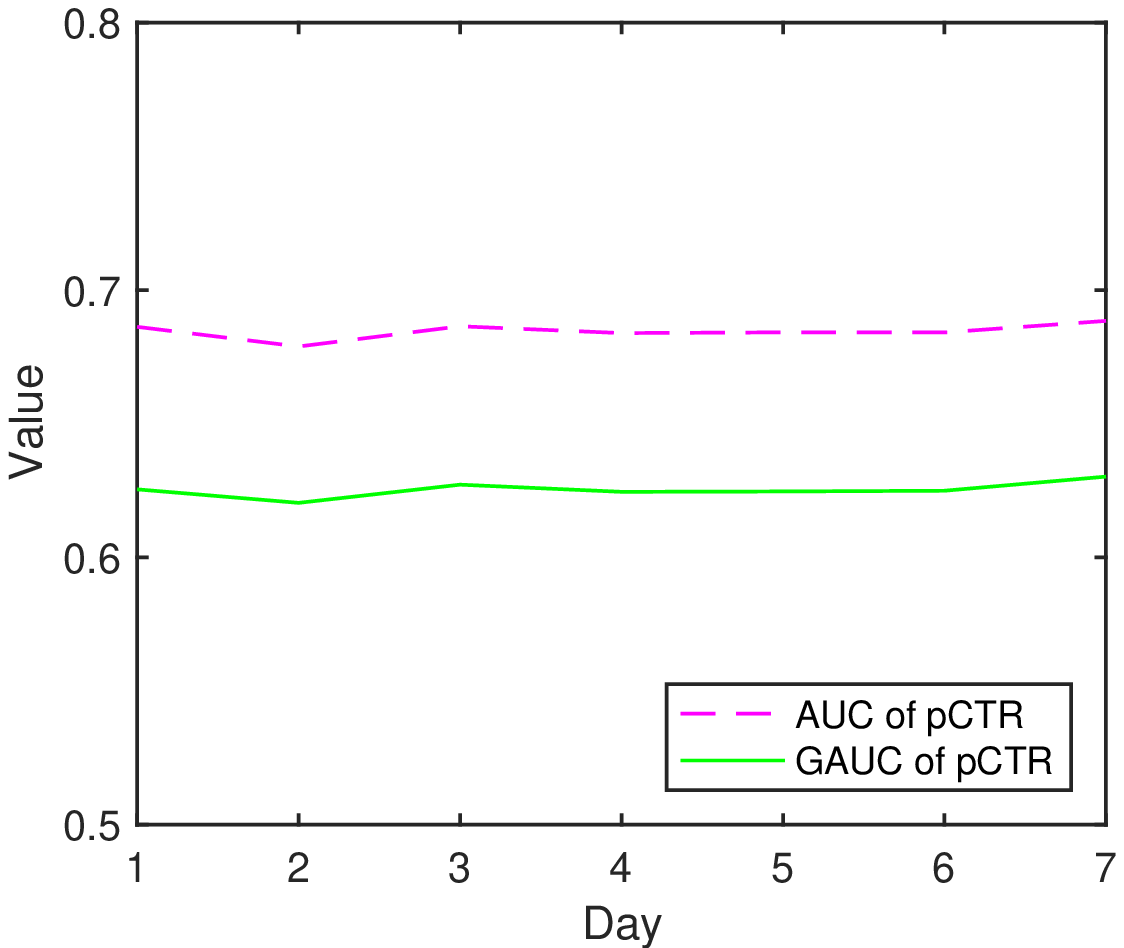}
        \label{fig:ctr_gauc}
    }
    \subfigure[CVR model]{
        \includegraphics[width=0.22\textwidth]{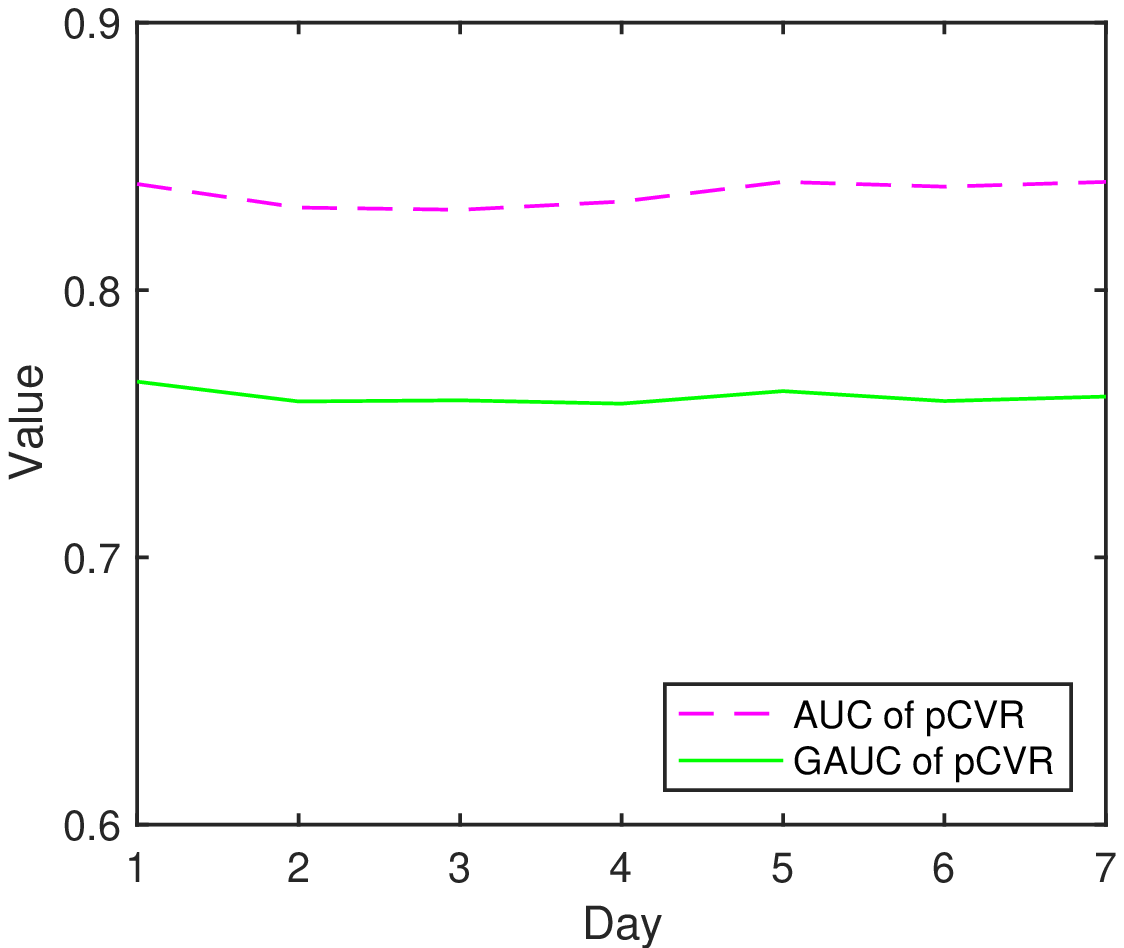}
        \label{fig:cvr_gauc}
    }
    \caption{The AUC and GAUC performance of CTR and CVR model in a 7 days period (from Jan 10, 2017 to Jan 16, 2017).}
    \label{fig:gauc}
\end{figure}

\begin{figure}[tb]
  \centering
  \includegraphics[width=0.6\columnwidth]{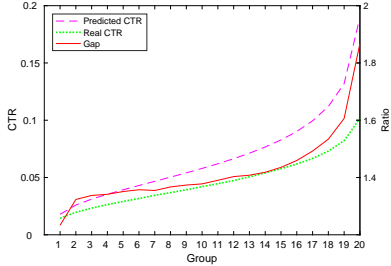}
  \caption{The gap between predicted and real CTR w.r.t. different pCTR level (from Jan 10, 2017 to Jan 16, 2017).}
  \label{fig:ctr_qcurve}
\end{figure}

Serving precise results are very important for prediction models. In tasks like CTR prediction, AUC is a widely used metric to measure model effectiveness. However, existing research \cite{cheng2016wide} shows that better AUC results in testing may bring worse performance in production. This also confused us in practice when tuning our prediction model.  We analyzed the problem and found that AUC metric doesn't treat different users and spots differently. For example, users who never click any ad or obscure ad spots would bring turbulence to AUC result towards a lower value. According to those facts and analysis, we proposed an AUC like metric, called Group AUC (GAUC) in Eq. \ref{Formula:gauc}. First, we aggregate all test data according to the user ($u$) and the particular position ($p$) of ad spot. Then, the AUC results are calculated in each single group (note that if there are all positive or negative samples in a group, we remove the group from the data). At last, we average these weighted AUC (weight $w_{(u, p)}$ is proportional to impression times or click times in the group) results in different groups and take the result as the GAUC value.


\begin{equation}
\label{Formula:gauc}
GAUC = \frac{\sum_{(u, p)}w_{(u, p)} * AUC_{(u, p)}}{\sum_{(u, p)}w_{(u, p)}}
\end{equation}
\paragraph{\textbf{CTR and CVR Model Performance}}
In Figure \ref{fig:gauc}, we give the AUC and GAUC performance of CTR and CVR prediction model in a 7 days period. The results show that the performance of daily models conducted by MLR algorithm are fairly stable. The CVR model has higher GAUC than CTR model, because there is less noises in the samples of CVR model.
In Figure \ref{fig:ctr_qcurve} and \ref{fig:cvr_qcurve}, we illustrate the ratio of predicted and real CTR, CVR values w.r.t. different predicted value levels. The results show that the predicted values of CTR are usually larger than the real ones. However, what matters more is the ordinal relation between different predicted CTR values in the proposed OCPC strategy.

The performance results show that the CTR and CVR prediction models, which are prerequisite of the proposed OCPC mechanism, are practicable.

\section{EXPERIMENTAL Results} \label{section:Experiment}
Satisfied by the above prediction model performance results, we are going to evaluate the effectiveness of the proposed OCPC method. The following experiments have two parts: offline simulation and online A/B test.

\subsection{Offline Simulation}
\label{section:offlineSimu}
In online advertising, it always take several days, or even several weeks, for new algorithms to take effect. Such long feedback time would put off the development and upgrading of new algorithm. To overcome the problem, we build an offline simulation platform to accelerate the validation of new ideas. Based on log data, the pre-view procedures can be restored perfectly. In other words, giving the same eligible ad list for each PV request, the auction-winning ads in simulation environment are the same with production environment. And for the coming post-view user behaviors, we use the predicted probability as a substitution of real clicks or conversions to estimate the real performance of different bid optimization strategies. For example, if an impressed ad's CTR prediction is $4\%$, then it will contribute $0.04$ to the total times of clicks. In the simulation, we use $20\%$ of all bidding records (which is about twenty million PVs) in \emph{Item CPC Ads} in Feb 11, 2017, and we compared 4 different bid optimization strategies:
\begin{itemize}
\item \textbf{Strategy 0} is the old strategy without bid optimization. And due to the sensibility and privacy of commercial data, the results of other strategies will be shown in comparison form contrast to this basic strategy. 

\item \textbf{Strategy 1} is a simple bid optimization strategy that takes the advertisers' view. Here we directly optimize $b_a^* = b_a * (1 + \sigma(\frac{p(c|u, a)}{E_u[p(c|u, a)]}, w) * r_a)$, where $\sigma(x, w) = \frac{x^w - 1}{x^w + 1}$ is a monotone increasing function (when $w > 0$) w.r.t. $x$, ranging in $(-1,1)$.

\item \textbf{Strategy 2} is our OCPC strategy that takes GMV pursuit of the traffic into account. The index $f(b_k^*) = pctr_k * b_k^* * (1 + \sigma(\frac{pcvr_k * v_k * \|A\|}{\sum_{i \in A}pcvr_i*v_i}, w) * r_a), w = 6$ and $r_a = 0.4$, where the implicit term ($pcvr_k * v_k$) could take effect to prompt GMV.

\item \textbf{Strategy 3} also attempts to promote GMV, but in the other way that directly sort eligible ads by descending order of $pctr*pcvr*bid$, without bid optimization.
\end{itemize}

\begin{figure}[h]
    \centering
    \subfigure[Curve of $\sigma(\cdot)$]{
        \includegraphics[width=0.22\textwidth]{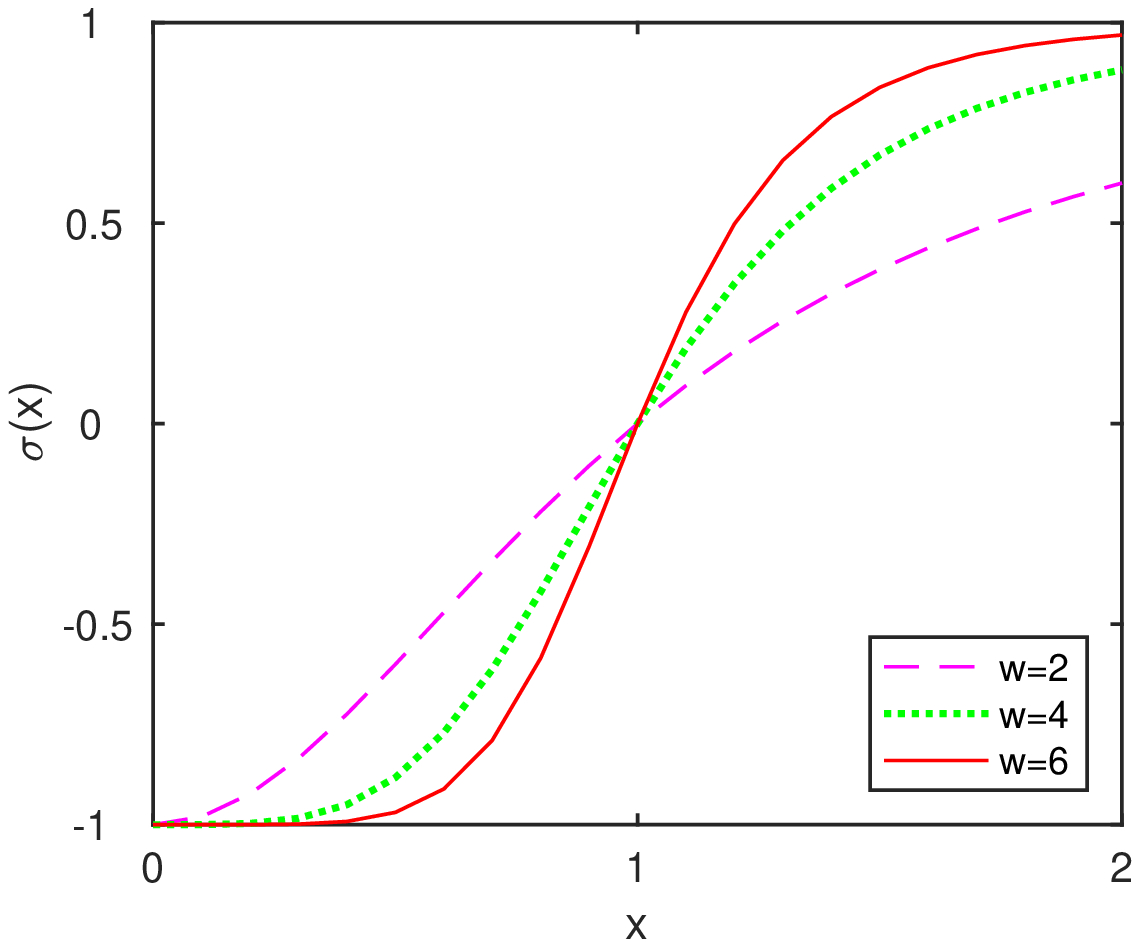}
        \label{fig:sigma}
    }
    \subfigure[Bid adjustment ratio]{
        \includegraphics[width=0.22\textwidth]{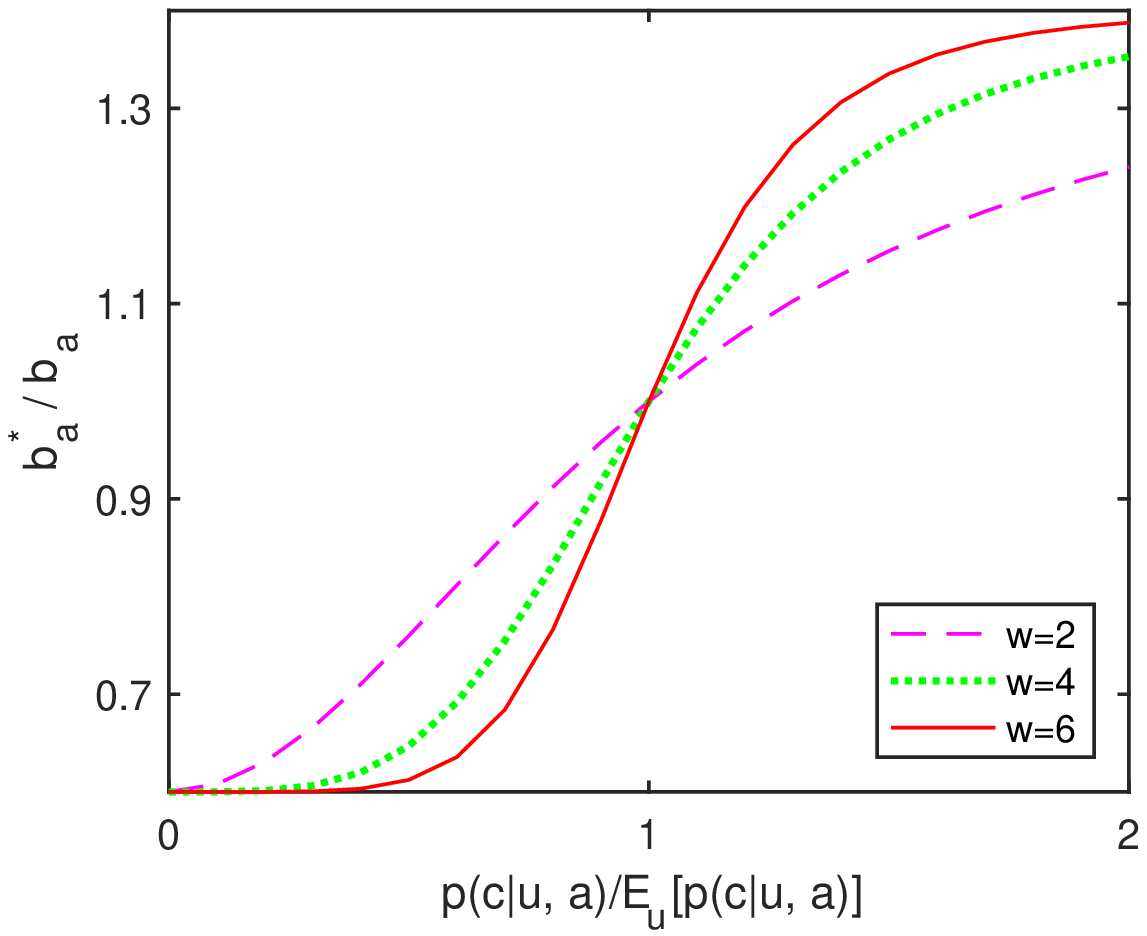}
        \label{fig:ratio}
    }
    \caption{The curve of $\sigma(\cdot)$ and the corresponding bid adjustment ratio of \textbf{Str} 1 when $r_a = 0.4$, w.r.t. different $w$.}
    \label{fig:sigma_curve}
\end{figure}

\textbf{Str} 1 is a straightforward strategy like the one proposed in \cite{perlich2012bid}, which attempts to optimize advertisers' ROI. The relationship between its bid optimization result and $p(c|u, a)$ is illustrated in Figure \ref{fig:sigma_curve}. \textbf{Str} 2 is the proposed OCPC strategy that has also considered Taobao's GMV pursuit. Using $pcvr$ and $v$ as the arguments of $\sigma(\cdot)$ in $f(\cdot)$, \textbf{Str} 2 tends to select those ads with high GMV estimation indirectly. \textbf{Str} 3 also attempts to promote GMV but in a new sorting mechanism out of eCPM.

Before giving the results, we will introduce some metrics in detail that we used to evaluate the performance of different bid optimization strategies. RPM is the indicator of advertising revenue per thousand impressions, which could measure the traffic liquidating efficiency of advertising platform. GMV per mille (GPM) is the gross merchandise volume per thousand impressions, which is related to advertisers' revenue and user experience of Taobao. ROI is to measure advertisers' return on investment. CTR, CVR and PPC are the average click-through rate, conversion rate and pay-per-click respectively.

\begin{table}[!hbp]
\small
\begin{tabular}{|c|c|c|c|c|c|c|}
\hline
& \textbf{RPM} & \textbf{GPM} & \textbf{ROI} & \textbf{CTR} & \textbf{CVR} & \textbf{PPC}\\
\hline
\textbf{Str 1} & -9.5\%& 8.8\%& 20.2\%& -0.5\%& 10.1\%& -7.8\%\\
\hline
\textbf{Str 2} & 5.6\%& 14.1\%& 8.1\%& -1.9\%& 14.9\%& 9.5\%\\
\hline
\textbf{Str 3} & -17.7\%& 23.6\%& 50.2\%& -8.6\%& 74.0\%& -9.8\%\\
\hline
\end{tabular}
\caption{Simulation results of different OCPC strategies when $r_a = 0.4$.}
\label{table:strategy}
\end{table}

In Table \ref{table:strategy}, we give the results of \textbf{Str} 1,2,3 against to \textbf{Str} 0. The parameter $w$ in adjustment function $\sigma(\cdot)$ is chosen by cross-validation and set to $2$ and $6$ for \textbf{Str} 1 and 2 respectively. \textbf{Str} 1 focuses on optimizing advertisers' ROI and cannot ensure better RPM. \textbf{Str} 3 can boost GPM by ranking with $pctr*pcvr*bid$,  however, it also pulls down RPM (because of pay per click (PPC) and CTR drop). Only the proposed OCPC strategy in \textbf{Str} 2 can achieve a tripartite win-win situation of GPM, ROI and RPM.


To measure the influence of different adjustment ranges $r_a$, we conduct an experiment with \textbf{Str} 2 (which outperforms the other strategies in the above experiment) and the results are shown in Table \ref{table:range}. Offline simulation results indicate that larger $r_a$ can bring better performance. And the increment of RPM is less than that of GPM, which results in a higher ROI lift when the adjustment range is large.

The results in Table \ref{table:strategy} and \ref{table:range} show that \textbf{Strategy 2} positively works in boosting overall GMV, and ROI constraint can protect advertisers' interests.
\begin{table}[!hbp]
\begin{tabular}{|c|c|c|c|c|c|c|c|}
\hline
\textbf{$r_a$}& \textbf{RPM} & \textbf{GPM} & \textbf{ROI} & \textbf{CTR} & \textbf{CVR} & \textbf{PPC}\\
\hline
0.2 & 4.2\%& 6.5\%& 2.2\%& -0.5\%& 6.5\% &5.5\%\\
\hline
0.3 & 5.2\%& 10.2\%& 4.8\%& -1.1\%& 10.4\%& 7.6\%\\
\hline
0.4 & 5.6\%& 14.1\%& 8.1\%& -1.9\%& 14.9\%& 8.1\%\\
\hline
0.5 & 5.5\%& 18.1\%& 11.9\%& -3.1\%& 19.9\%& 11.2\%\\
\hline
\end{tabular}
\caption{Simulation results of \textbf{Str} 2 w.r.t. different $r_a$.}
\label{table:range}
\end{table}
\paragraph{\textbf{Campaign Results}}
Besides the overall performance, we also simulate the performance of particular campaigns under \textbf{Str} 2, to ensure that the proposed strategy can improve each single campaign's advertising effect. The results of 10 campaigns with largest cost in simulation are shown in Table \ref{table:campaign}. The metric named "Cost" is the total payment for advertising. An interesting observation is that seven campaigns' GPM increases while their PV drops at the same time, which means that they win less poor quality opportunities with OCPC mechanism. In addition, eight in ten campaigns' ROI is improved, which shows that the ROI constraint truly work for separate campaigns. The ROIs of campaign 3 and 8 drop slightly, because they compete more PVs.


\begin{table}[!hbp]
\small
\begin{tabular}{|c|c|c|c|c|c|c|c|}
\hline
& \textbf{GMV} & \textbf{Cost} & \textbf{PV} & \textbf{GPM} & \textbf{ROI} \\
\hline
\textbf{Camp 1} & -0.9\%& -17.5\%& -16.2\%& 18.2\%& 20.1\%\\
\hline
\textbf{Camp 2} & -7.7\%& -27.9\%& -26.8\%& 26.2\%& 28.1\%\\
\hline
\textbf{Camp 3} & 2.5\%& 9.2\%& 2.6\%& -0.1\%& -6.2\%\\
\hline
\textbf{Camp 4} & 23.0\%& 9.2\%& 0.4\%& 22.6\%& 12.7\%\\
\hline
\textbf{Camp 5} & -13.1\%& -23.8\%& -22.0\%& 11.4\%& 14.0\%\\
\hline
\textbf{Camp 6} & 0.0\%& -4.0\%& -10.3\%& 11.5\%& 4.1\%\\
\hline
\textbf{Camp 7} & -5.0\%& -8.0\%& -9.7\%& 5.2\%& 3.2\%\\
\hline
\textbf{Camp 8} & 64.6\%& 65.7\%& 49.5\%& 10.1\%& -0.6\%\\
\hline
\textbf{Camp 9} & -19.2\%& -30.4\%& -28.5\%& 13.1\%& 16.1\%\\
\hline
\textbf{Camp 10} & -4.2\%& -5.3\%& -8.6\%& 4.9\%& 1.2\%\\
\hline
\end{tabular}
\caption{Simulation results of \textbf{Str} 2 w.r.t. different campaigns.}
\label{table:campaign}
\end{table}

\subsection{Online Results of OCPC Strategy 2}
After offline simulation and online mini flow A/B test in experimental environment, we finally decide to deploy aforementioned \textbf{Str} 2 in production. Meanwhile, \textbf{Str} 0 is reserved as a contrast test. In this section, we are going to study the online performance of proposed OCPC strategy in \emph{Item CPC Ads}. And other results of different traffic pursuits and scenarios are also shown to prove the effectiveness and generality of OCPC mechanism.

In Table \ref{table:flow}, we give the experimental results of \textbf{Str} 2 with $30\%$ of whole production traffic, and the benchmark \textbf{Str} 0 also has $30\%$ traffic. Users are allocated to each strategy randomly, while all ad campaigns exist in both strategies. Note that we have about ninety million PVs every day in \emph{Item CPC Ads}. The results prove stable improvement of the proposed bid optimization strategy. Advertisers' interests (indicated by ROI), platform's revenue (indicated by RPM) and overall GPM achieve a tripartite win-win situation.



\begin{table}[!hbp]
\small
\begin{tabular}{|c|c|c|c|c|c|c|}
\hline
& \textbf{RPM} & \textbf{GPM} & \textbf{ROI} & \textbf{CTR} & \textbf{CVR}\\
\hline
\textbf{\% Improved} & 6.6\%& 8.9\%& 2.1\%& -1.3\%& 5.2\%\\
\hline 
\end{tabular}
\caption{Online experimental results of \textbf{Str} 2 under $30\%$ of whole production traffic ($r_a = 0.4$, from Aug 23, 2016 to Aug 29, 2016).}
\label{table:flow}
\end{table}

After giving the results about overall performance, we do other experiments (from Sep 8, 2016 to Sep 14, 2016) to verify the effectiveness of the \textbf{Str} 2 further, to find out whether it benefit to most separate advertisers and the advertising platform in the long term. 

\paragraph{\textbf{Performance in Advertisers' View}}
Firstly, we analyze the performance for each separate campaign. Campaigns with at least $5$ conversions in a week are included. In Table \ref{table:advertiserView}, we give the proportion data of ad campaigns whose ad performance is improved. In all campaigns with more than $5$ conversions in a week, $67\%$ campaigns get GPM and ROI improvement at the same time. And $24\%$ campaigns are in the so called \textbf{quantity and quality exchange} situation: their PV increment is larger than the ROI drop. We say that it's also acceptable for some advertisers, because PV increment might lead those secondary impressions to a campaign and lower the ROI. However, more impressions could also bring more conversions.

\begin{table}[!hbp]
\begin{tabular}{|c|c|c|c|c|c|c|}
\hline
& \textbf{$\%$ Campaigns}\\
\hline
\textbf{GPM and ROI are improved} & 67\%\\
\hline
\textbf{Quantity and quality exchange} & 24\%\\
\hline 
\end{tabular}
\caption{The proportion of ad campaigns whose performance is improved. Here we choose campaigns with more than $5$ conversions in the experiment.}
\label{table:advertiserView}
\end{table}

\begin{figure}[tb]
  \centering
  \includegraphics[width=0.7\columnwidth]{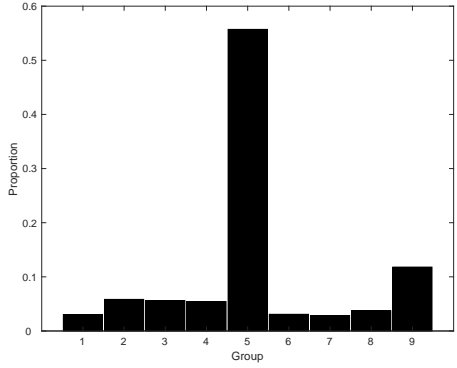}
  \caption{The proportion of different ratios of optimized bid price and determined bid.}
  \label{fig:tuneRatio}
\end{figure}

With OCPC mechanism, advertisers might also be concerned about what the optimized bid prices actually are. In Figure \ref{fig:tuneRatio}, we illustrate the numerical relation between optimized bid price $b_a^*$ and determined bid $b_a$ for those displayed ads in Feb 19, 2017. We divide those bidding records into 9 groups, according to their value of $b_a^* / b_a$ (ranging from $1-r_a$ to $1 + r_a$). From the results, we can see that more than half impressions belong to group 5, the middle group which includes records with $b_a^* = b_a$. It's a reasonable observation, because the bid optimization upper bound for those low quality traffics is set to $b_a$ according to Eq. \ref{Formula:upper_bound}, and the proposed ranking algorithm prefer to adopt the upper bound.

\paragraph{\textbf{Performance in Platform's View}}
Standing by the platform side, merely focusing on the overall RPM, GPM and ROI results is far from enough. In Taobao advertising system, ad items are from variant kinds of categories, e.g., women's dress, furniture or digital product. For each category, it has an inherent CVR or ROI level. There exists the probability that the overall improvements of GMV or ROI come from the traffic shifting between different categories, which is not good in the long run. Thus, we give an experimental result to capture the traffic shifting.

\begin{figure}[tb]
  \centering
  \includegraphics[width=0.7\columnwidth]{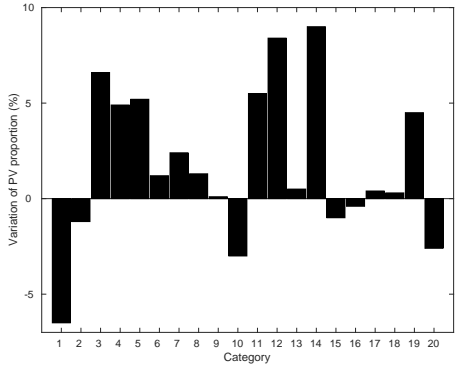}
  \caption{The variation of PV proportion of top 20 categories (ranked by category's total advertising cost).}
  \label{fig:shift}
\end{figure}

The results of variation of PV proportion are given in Figure \ref{fig:shift}. PV proportion of a category is the ratio of the category's PV and the total PV in an experiment bucket. The results suggest that the traffic shifting is not too obvious, with all the variations are within $\pm 10\%$ (note that the PV proportion might change whether the algorithms are different, in different buckets).

Analogy to the advertisers' view, we also do experiments to show the performance in categories' view. Results in Table \ref{table:platform} suggest that $17\%$ categories (with $62\%$ PVs) get GPM and ROI improvement at the same time.
\begin{table}[!hbp]
\small
\begin{tabular}{|c|c|c|c|c|c|c|}
\hline
& \textbf{$\%$ Category} & \textbf{$\%$ PV}\\
\hline
\textbf{GPM and ROI are improved} & 17\%& 62\%\\
\hline
\textbf{GPM is improved} & 27\%& 21\%\\
\hline 
\textbf{Quantity and quality exchange} & 30\%& 12\%\\
\hline
\end{tabular}
\caption{The proportion of categories whose performance is improved, and the corresponding PV proportion of categories.}
\label{table:platform}
\end{table}

The results in platform's and advertisers' view prove that the OCPC algorithm has the capability to hand out suitable opportunities to different ads, which can improve comprehensive utilization effect of advertising traffic. And all the above results prove that OCPC takes significant effect for both Taobao advertising platform and the advertisers.

\subsection{Online Performance in Other Scenarios}
As mentioned in Section \ref{section:optScope}, advertisers could have different pursuits. Before Double Eleven, Taobao sellers are more concerning about the quantity of goods added to users' shopping cart. We do experiments with \textbf{Str} 2 using different $f(\cdot)$ before 2016's Double Eleven event. Using the predicted probability of adding to shopping cart (predicted ASR), the index function $f(k, b_k^*) = pctr_k * b_k^* * (1 + \sigma(\frac{pasr_k * \|A\|}{\sum_{i \in A}pasr_k}, w) * r_a)$, $w = 6$ and $r_a = 0.4$.

In Table \ref{table:pursuit}, we give the results of OCPC strategy which helps prompt ASR. From the results, we can observe that ASR has been improved $15.6\%$ (compared to \textbf{Str} 2 with aforementioned $f(\cdot)$ in Section \ref{section:offlineSimu}).

\begin{table}[!hbp]
\small
\begin{tabular}{|c|c|c|c|c|c|c|}
\hline
& \textbf{GPM} & \textbf{RPM} & \textbf{CTR} & \textbf{CVR} & \textbf{ASR}\\
\hline
\textbf{$\%$ Improved} & 0.3\%& -6.1\%& -2.9\% & 21.1\%& 15.6\%\\
\hline
\end{tabular}
\caption{The online results about prompting the probability of adding to shopping cart (from Oct 30, 2016 to Nov 10, 2016, $5\%$ production flow).}
\label{table:pursuit}
\end{table}

Besides, we give the results of \textbf{Str} 2 in \emph{Banner CPC Ads}, with $f(k, b_k^*) = pctr_k * b_k * (1 + \sigma(\frac{pcvr_k * \|A\|}{\sum_{i \in A}pcvr_i}, w) * r_a)$ instead, $w = 6$ and $r_a = 0.4$, in Table \ref{table:banner}. Note that we remove $v_a$ term from $f(\cdot)$, because there are store campaigns in which PPBs of different items vary a lot. The results suggest large CVR and GPM improvement.

\begin{table}[!hbp]
\small
\begin{tabular}{|c|c|c|c|c|c|c|}
\hline
& \textbf{GPM} & \textbf{RPM} & \textbf{ROI} & \textbf{CTR} & \textbf{CVR}\\
\hline
\textbf{$\%$ Improved} & 15.7\%& 3.6\%& 11.7\% & -0.6\%& 19\%\\
\hline
\end{tabular}
\caption{The online results in \emph{Banner Ads} (from Jan 13, 2017 to Jan 15, 2017, $30\%$ production flow).}
\label{table:banner}
\end{table}

Above experiments show that the OCPC mechanism could act like a general framework to handle different problems, no matter what the pursuits and scenarios are.


\section{Conclusion}\label{section:Conclusion}
We introduce a number of important features of Taobao display advertising system, and elaborate on two key ad formats, i.e., banner and item ads. By analyzing the ecological characteristics and comparing with other methods, we use the most suitable pricing method, i.e., CPC in the involved ads formats. We showcase our system architecture and ads serving process, based on which we analyzed the shortcomings of the traditional CPC method and propose OCPC algorithm to reconcile the demands of advertisers, platform ecological indices and platform revenue. We characterize the optimization objectives mathematically and give detailed algorithms with other relative technical details such as prediction models, calibration, and algorithm complexity analysis. Holding eCPM sorting mechanism, our proposed OCPC strategy benefits to not only advertisers, but also other indices including eCPM itself, by bid optimization. In Taobao display advertising platform, OCPC has been automatic applied in the whole mobile production traffic of \emph{Item CPC Ads}, and can also be chosen to apply by advertisers in their own \emph{Banner CPC Ads} traffic.

\bibliographystyle{ACM-Reference-Format}
\bibliography{OCPC} 

\end{document}